\documentclass[prl,amsmath,amssymb,twocolumn]{revtex4}

\usepackage{graphicx}
\usepackage{subfigure}
\usepackage{adjustbox}
\usepackage{bm}
\usepackage{color}
\usepackage{braket}
\usepackage{standalone}
\usepackage{multirow}
\usepackage{tikz}
\usepackage{mathrsfs}
\usepackage[colorlinks,bookmarks=true,citecolor=blue,linkcolor=red,urlcolor=blue]{hyperref}

\begin{document}
\title{Lattice models with exactly solvable topological hinge and corner states}

\author{Flore K. Kunst$^{1}$, Guido van Miert$^{2}$ and Emil J. Bergholtz$^{1}$}

\affiliation{$^1$ Department of Physics, Stockholm University, AlbaNova University Center, 106 91 Stockholm, Sweden
\\$^2$ Institute for Theoretical Physics, Centre for Extreme Matter and Emergent Phenomena, Utrecht University, Princetonplein 5, 3584 CC Utrecht, The Netherlands}
\date{\today}

\begin{abstract}
We devise a generic recipe for constructing $D$-dimensional lattice models whose $d$-dimensional boundary states, located on surfaces, hinges, corners, and so forth, can be obtained exactly. The solvability is rooted in the underlying lattice structure and as such does not depend on fine tuning, allowing us to track their evolution throughout various phases and across phase transitions. Most saliently, our models provide ``boundary solvable" examples of the recently introduced higher-order topological phases. We apply our general approach to breathing and anisotropic kagome and pyrochlore lattices for which we obtain exact corner eigenstates, and to periodically driven two-dimensional models as well as to three-dimensional lattices where we present exact solutions corresponding to one-dimensional chiral states at the hinges of the lattice. We relate the higher-order topological nature of these models to reflection symmetries in combination with their provenance from lower-dimensional conventional topological phases.
\end{abstract}

\maketitle

{\it Introduction.}
Topological phases of matter have been a central research topic in condensed matter physics over the past decade. Famous examples are Chern insulators \cite{tknn, haldane, changzhangfengshenzhang, jotzumesserdesbuquoislebratuehlingergreifesslinger} and $\mathbb{Z}_2$ insulators \cite{kaneandmele, kaneandmeletwo, bernevigzhang, bernevighugheszhang, koenigwiedmannbruenerothbuhmannmolenkamp}, which are lattice realizations of the integer \cite{klitzingdordapepper} and spin quantum Hall effect, respectively, as well as three-dimensional (3D) Weyl semimetals \cite{volovik,murakami, wanturnerbishwanathsavrasov, burkovbalents, luwangyeranfujoannopoulossoljacic, xubelopolskialidoustetal, lvwengwantmiaoetal}. The main characteristic of these phases is the presence of robust, topological boundary states, which appear in the form of chiral edge states, helical edge states, and chiral Fermi arcs for the above-mentioned models, respectively. Very recently, higher-order topological phases with concomitant robust corner and hinge states have been proposed \cite{benalcazarbernevighughes} and triggered an avalanche of followup works \cite{parameswaranwan,langhehnpentrifuoppenbrouwer,linhughes, songfangfang, benalcazarbernevighughesagain, schindlercookvergnio, ezawapap, xuxuewan, imhofbergerbayerbrehmmolenkampkiesslingschindlerleegreiterneupertthomale, garciaperissstrunkbilallarsenvillanuevahuber, petersonbenalcazarhughesbahl}.

In this Rapid Communication, we devise an approach to engineering lattice models that possess corner and hinge states whose wave functions can be readily obtained exactly. The exact solutions, which generalize the work on conventional topological edge and surface states in Refs.~\cite{kunsttrescherbergholtz,bergholtz2015}, can be obtained independently of tight-binding parameters while crucially depending on the underlying lattice structure, making them remarkably robust and allowing a deeper microscopic understanding of these intriguing states of matter. After providing a generic recipe for devising $D$-dimensional models with solvable $d$-dimensional boundary states, we apply this to two-dimensional (2D) breathing kagome and three-dimensional breathing pyrochlore models possessing topologically protected zero-dimensional corner states \cite{ezawapap}. Moreover, we introduce several three-dimensional models that support one-dimensional (1D) hinge states that are chiral, tunable, and readily deduced from local destructive interference arising from the underlying lattice geometry.  

\begin{figure}[t]
{\includegraphics[width=1\linewidth]{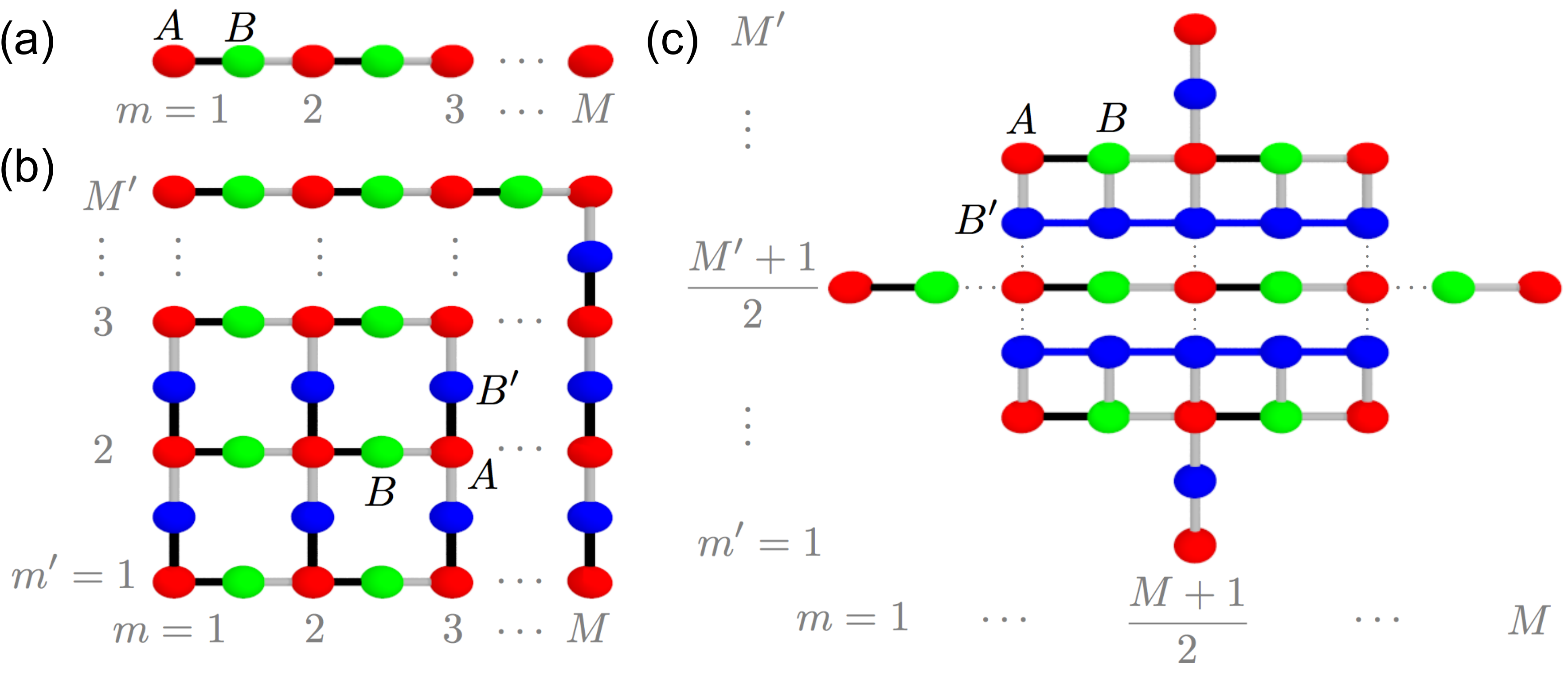}}
\caption{Examples of lattice structures with solvable corner states. The $A$, $B$, and $B'$ lattices are colored red, green, and blue, respectively, while the black and gray lines indicate inequivalent hoppings between the $A$ and $B$ or $A$ and $B'$ lattices.}
\label{fig_general_structure}
\end{figure}

{\it Setup.} We consider tight-binding models, described by local hopping Hamiltonians, $H=\sum_{i,j}t_{ij}c^\dagger_ic_j$, on $D$-dimensional lattices with open boundary conditions in $D-d$ dimensions as illustrated in Fig.~\ref{fig_general_structure} for $D=1$ [Fig.~\ref{fig_general_structure}(a)] and $(D=2)$-dimensional [Figs.~\ref{fig_general_structure}(b) and \ref{fig_general_structure}(c)] example lattices whose $(d=0)$-dimensional boundaries, i.e., their end and corner states, respectively, are solvable. The lattices are composed of three different sublattices labeled $A$ (red), $B$ (green), and $B'$ (blue), and we allow generic hopping terms as long as the $A$ lattices are only connected via direct hopping to adjacent $B$ and $B'$ lattices. The lattice terminates with $A$ lattices at each of its corners, which is crucial for the solvability of the boundary states. 

If we now consider a linear combination of eigenstates, $c^\dagger_{A_i,m,m'}\ket{0}$, of each $A$ lattice with each the same energy $E_i$, this can also be an eigenstate (with energy $E_i$) of the full system if---and only if---the local eigenstates interfere destructively, resulting in a zero amplitude at each of the $B$ and $B'$ lattices. Solutions of this form are given by 
\begin{equation}
\ket{\Psi_i} = \mathcal{N}_i \sum_{m=1}^{M} \sum_{m'=1}^{M'} r_i^{m} {r'}_i^{m'} c^\dagger_{A_i,m,m'}\ket{0}, \label{eqexactsoldc}
\end{equation}
where $\mathcal{N}_i$ is a normalization factor, $c^\dagger_{A_i,m,m'}$ creates a local $A$-lattice eigenstate $i$, and $r_i,r_i'$ are complex numbers whose explicit form is determined to ensure destructive interference such that amplitudes on the $B$ and $B'$ lattices vanish. 

A particularly simple and important special case is the Su-Schrieffer-Heeger (SSH) chain \cite{ssh} featuring a single orbital in both $A$ and $B$ and with nearest-neighbor hopping amplitudes $t_1$ [black lines in Fig. \ref{fig_general_structure}(a)] and $t_2$ (gray): An exact end state is then given by Eq.~(\ref{eqexactsoldc}) with $M'=1$ and $r=-t_1/t_2$ implying that the state is exponentially localized to the left ($m=1$) end for $|t_1|<|t_2|$ and to the right ($m=M$) end when $|t_1|>|t_2|$. Notably, the gap in the bulk Hamiltonian, $\mathcal{H}_k = [t_1 + t_2 \, \textrm{cos}(k)]\sigma_x + t_2 \, \textrm{sin} (k) \sigma_y$, closes at $k = \pi \, \textrm{mod} \, 2 \pi$ ($k = 0 \, \textrm{mod} \, 2 \pi$) when $t_1 = +(-) t_2$, i.e., $|r|=1$, such that the delocalization of the end mode is accompanied by a topological phase transition.

In fact, in the case of just a single orbital in $B$ and $B'$, consistent solutions for $r$ and $r'$ always exist. In the more general case, solutions rely on discrete symmetries, as in, e.g., Fig.~\ref{fig_general_structure}(c), where a mirror symmetry is sufficient to guarantee their existence. This thus allows us to find a set of eigenstates of a system of arbitrary size at the prize of only diagonalizing the Hamiltonian of an extremely small subsystem, i.e., that of the single $A$ unit cell. Moreover, it is clear that Eq.~(\ref{eqexactsoldc}) generically describes corner states, which are exponentially localized to either of the four corners depending on the absolute values of $r_i$ and $r_i'$, while they are delocalized only for fine-tuned parameter values with diverging localization lengths, $\xi_i=1/\log{|r_i|}$ and/or $\xi'_i=1/\log{|r'_i|}$.

\begin{figure}[t]
  \centering
  {\includegraphics[width=\linewidth]{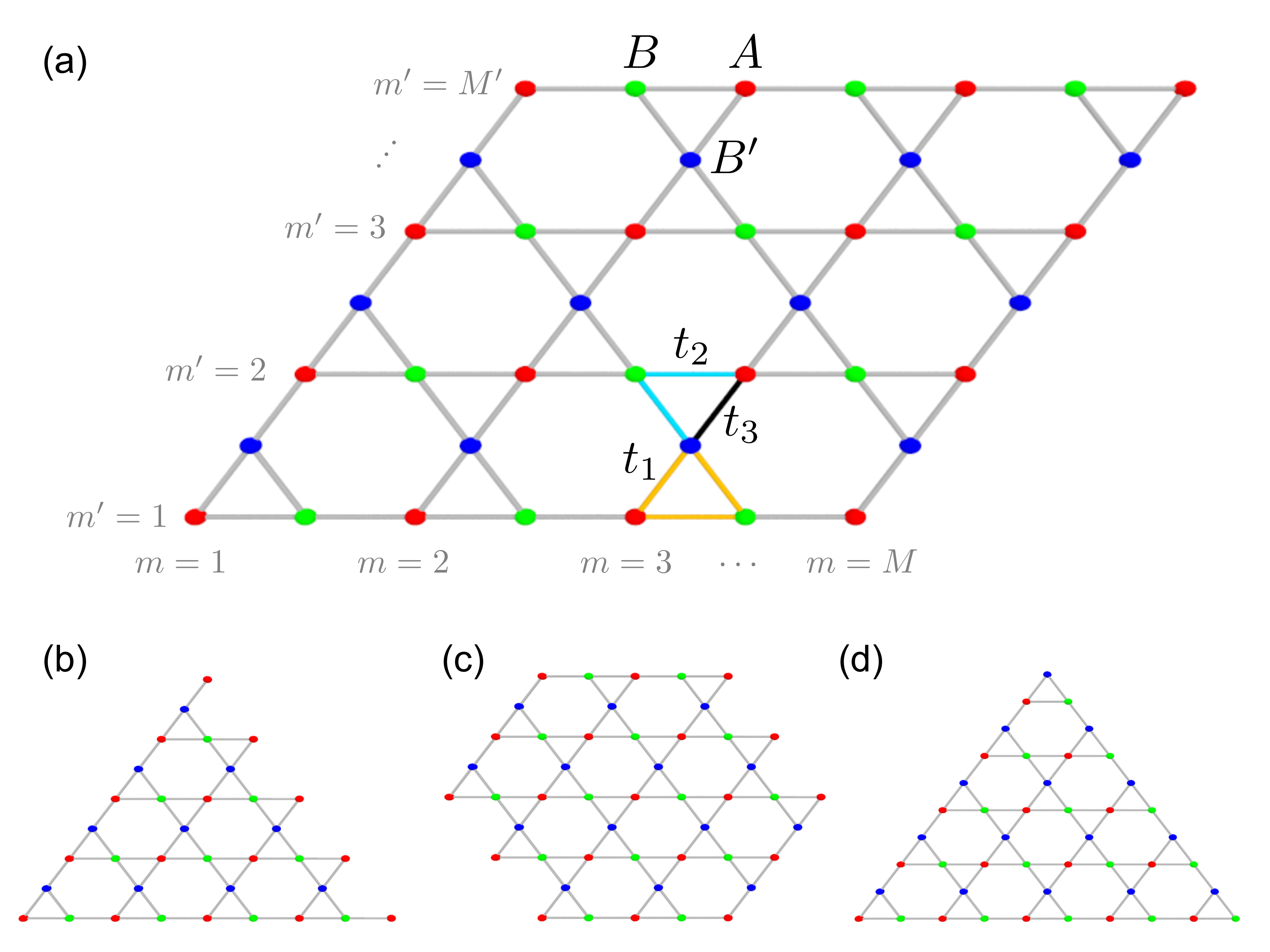}}
\caption{Breathing kagome lattice in the shape of (a) a rhombus, (b) a triangle with bearded edge, (c) a hexagon with two bearded edges, and (d) a triangle with smooth edges. The $A$, $B$, and $B'$ lattices are shown in red, green, and blue, respectively. In (a) the hopping parameters are explicitly shown on one up-triangle and one down-triangle. The orange, blue, and black bonds have hopping parameters $t_1$, $t_2$, and $t_3$, respectively. Equation (\ref{kagomestates}) is an exact eigenstate for (a), (b), and (c) while it also provides key insight into (d) albeit not an exact finite-size eigenstate for the smooth triangle.}
\label{fig_breathing_kagome_lattice}
\end{figure}

The generalization to any $D$ and $d$ is simple and direct; $d$ is increased by introducing further internal periodic variables, $\mathbf k = (k_1, \ldots, k_d)$, i.e., via dimensional extension, while the codimension $D-d$ is set by the number of intermediate $B,B',B'',\ldots,B^{(D-d-1)}$ lattices in the full lattice and as such can be engineered to assume any integer value leading to generic eigenstate solutions of the form 
\begin{align}
\ket{\Psi_i (\mathbf k)}=& \mathcal{N}_i (\mathbf k) \nonumber\\
& \times \sum_{\{m^{(s)}\}}\!\! \left[\!\prod_{s=0}^{D-d-1}\!\left[r_i^{m^{(s)}}(\mathbf k)\right]^{m^{(s)}}\right]\!\!c^\dagger_{A_i,\mathbf k, \{m^{(s)}\}}\! \ket{0}\!,
\label{eqexactsolgeneral}
\end{align}
where the expressions $r_i^{(s)}(\mathbf k)$ now acquire an explicit momentum dependence $\mathbf k$, opening the possibility of boundary switching the state $\ket{\Psi_i (\mathbf k)}$ as a function of $\mathbf k$. The sum over the set $\{m^{(s)}\}=\{m,m',\ldots, m^{(D-d-1)}\}$ runs over the entire lattice.

\begin{figure}[h]
  \centering
  {\includegraphics[width=1\linewidth]{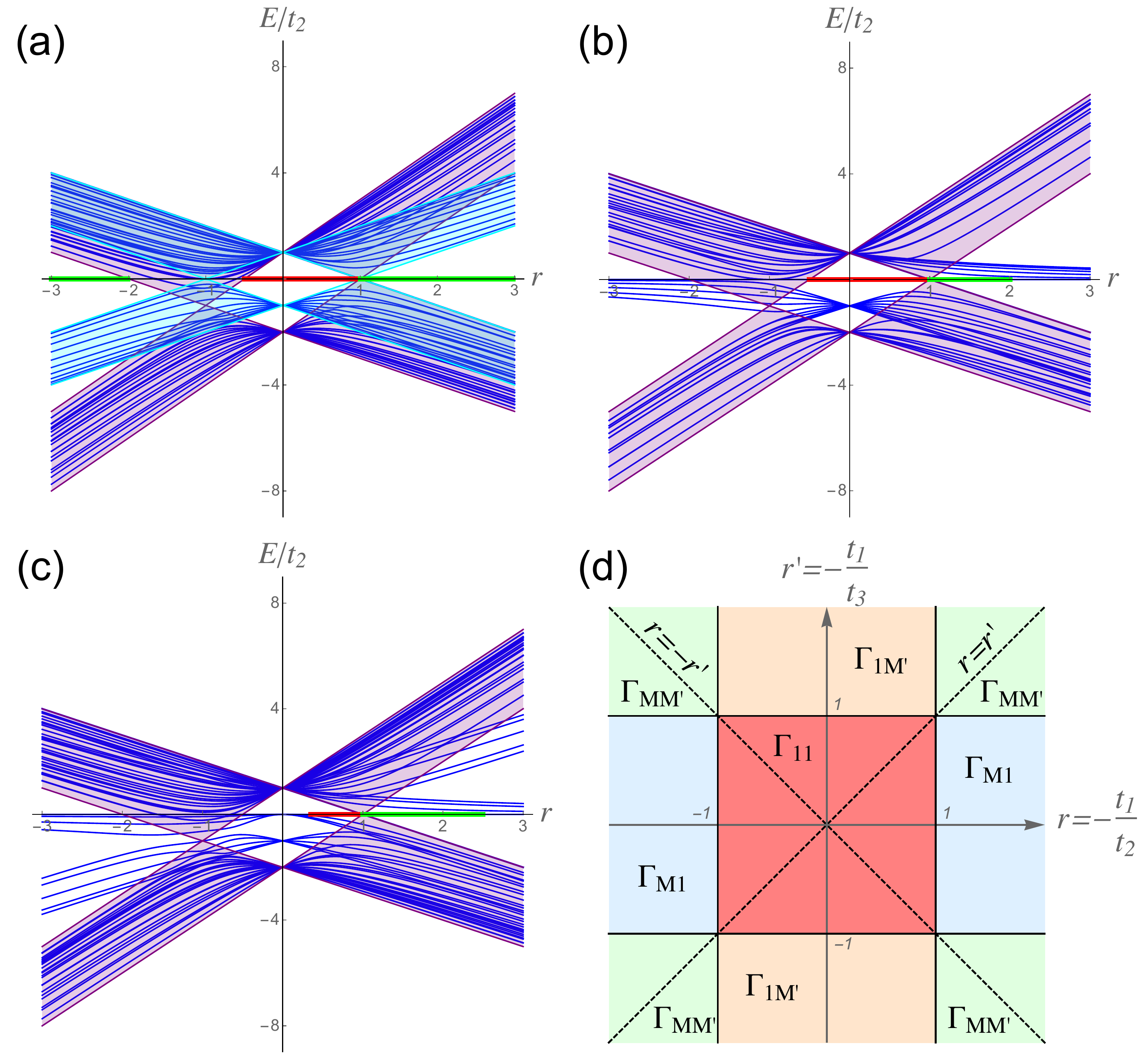}}
\caption{Energy spectra for the breathing kagome lattice in the shape of (a) a rhombus, (b) a bearded triangle, and (c) a bearded hexagon with $r = r'$ and (a), (b) ${\rm max}(M) = {\rm max}(M') = 6$, and (c) ${\rm min}(M) = {\rm min}(M') = 3$, ${\rm max}(M) = 6$, and ${\rm max}(M') = 7$. The bands in the purple shaded area originate from the bulk of the system, whereas the other blue bands originate from the edges. All edges of the rhombus in Fig.~\ref{fig_breathing_kagome_lattice}(a) have a structure corresponding to Fig.~\ref{fig_general_structure}(a) such that we know their spectrum, and the bands originating from these edges are shown explicitly shown in cyan in (a). Notably, the bands originating from the bulk and the edges overlap. The exact corner solutions are at zero energy and are colored in accordance with the phase diagram in (d), where $\Gamma_{mm'}$ indicate the localization of the state to the corresponding corner/bearded edge.}
\label{fig_breathing_kagome_lattice_plots}
\end{figure}

{\it Breathing kagome corner states.}
Having provided the generic strategy, we now proceed to show that the above formalism provides exact zero-energy solutions corresponding to corner states in the breathing kagome lattice. This can be seen as a natural generalization of the SSH model to two dimensions and as such it inherits, {\it mutatis mutandis}, its key properties. The pertinent Hamiltonian includes real nearest-neighbor hopping terms only, as shown in yellow, blue, and black in Fig.~\ref{fig_breathing_kagome_lattice}(a), and reduces to the model studied in Refs.~\cite{ezawapap, xuxuewan} for $t_2 = t_3$. The nonzero terms in the Bloch Hamiltonian read $\mathcal{H}_{AB} = -t_1 - t_2 \, e^{i k_x}$, $\mathcal{H}_{AB'} = -t_1 - t_3 \, e^{i (k_x/2 + \sqrt{3} k_y/2)}$, $\mathcal{H}_{BB'} = -t_1 - t_2 \, e^{-i (k_x/2 - \sqrt{3} k_y/2)}$, supplemented by their Hermitian conjugates. Whereas corner states were found numerically in Refs.~\cite{ezawapap, xuxuewan}, we find that their exact form on the rhombus, bearded triangle, and bearded hexagon geometries shown in Figs.~\ref{fig_breathing_kagome_lattice}(a)-\ref{fig_breathing_kagome_lattice}(c) is 
\begin{equation}
\ket{\Psi_{{\rm corner}}} = \mathcal{N} \sum_{m,m'}\left(-\frac{t_1}{t_2}\right)^{m} \left(-\frac{t_1}{t_3}\right)^{m'} c^\dagger_{A,m,m'}\ket{0}, \label{kagomestates}
\end{equation}
where $r = -t_1/t_2,\ r' = -t_1/t_3$ in Eq.~(\ref{eqexactsoldc}) solves the Schr\"{o}dinger equation for the system with open boundaries by virtue of destructive interference such that the amplitudes on the $B$ and $B'$ lattices vanish. While this is not an exact solution for the triangle geometry in Fig.~\ref{fig_breathing_kagome_lattice}(d), Eq.~(\ref{kagomestates}) provides crucial insight also into this case by noting that each corner maps locally onto the lower left corner in Fig.~\ref{fig_breathing_kagome_lattice}(a): In the regime $|r|, |r'|<1$ there are three near-zero-energy states which are, in the large system limit, captured by Eq.~(\ref{kagomestates}) and two duplicates thereof rotated to the other two corners. Whenever $|r|>1$ or $|r'|>1$, there are, however, no corner states in this geometry, consistent with the absence of a state localized to the lower left corner in Fig.~\ref{fig_breathing_kagome_lattice}(a), and signaling a ``trivial" phase.

Figure \ref{fig_breathing_kagome_lattice_plots} shows the energy spectrum for the rhombus [Fig.~\ref{fig_breathing_kagome_lattice_plots}(a)], bearded triangle [Fig.~\ref{fig_breathing_kagome_lattice_plots}(b)], and bearded hexagon [Fig.~\ref{fig_breathing_kagome_lattice_plots}(c)] geometries, all with $r = r'$, and the phase diagram in [Fig.~\ref{fig_breathing_kagome_lattice_plots}(d)]. The zero-energy states correspond to the exact solution in Eq.~(\ref{kagomestates}), and it localizes to the corners $\{m,m'\}=\{1,1\}$ and $\{m,m'\}=\{M,M'\}$ in the red and green regions, respectively, where the accessibility of the four different localization sectors $\Gamma_{mm'}$ depends on the values of $r$ and $r'$ and is shown in the phase diagram.  

To understand the phase diagram it is key to realize that edge and bulk bands necessarily attach to the zero-energy mode in the large system limit whenever $|r|=1$ and/or $|r'|=1$ \cite{kunsttrescherbergholtz}. This is indeed a necessary condition enabling the state to ``switch" corners, which is indeed observed in Figs.~\ref{fig_breathing_kagome_lattice_plots}(a)-\ref{fig_breathing_kagome_lattice_plots}(c) and shown in the phase diagram in Fig.~\ref{fig_breathing_kagome_lattice_plots}(d). We note, moreover, that the bulk/edge attachment at zero energy can extend beyond $|r|=1$ and/or $|r'|=1$: While the zero-energy mode is isolated in the bulk gap in parts of the sectors $\Gamma_{11}$ and $\Gamma_{MM'}$, the sectors $\Gamma_{1M'}$ and $\Gamma_{M1}$ are entirely characterized by the attachment of noncorner states to zero energy, i.e., they correspond to gapless phases.

\begin{figure}[h]
  \centering
  {\includegraphics[width=1\linewidth]{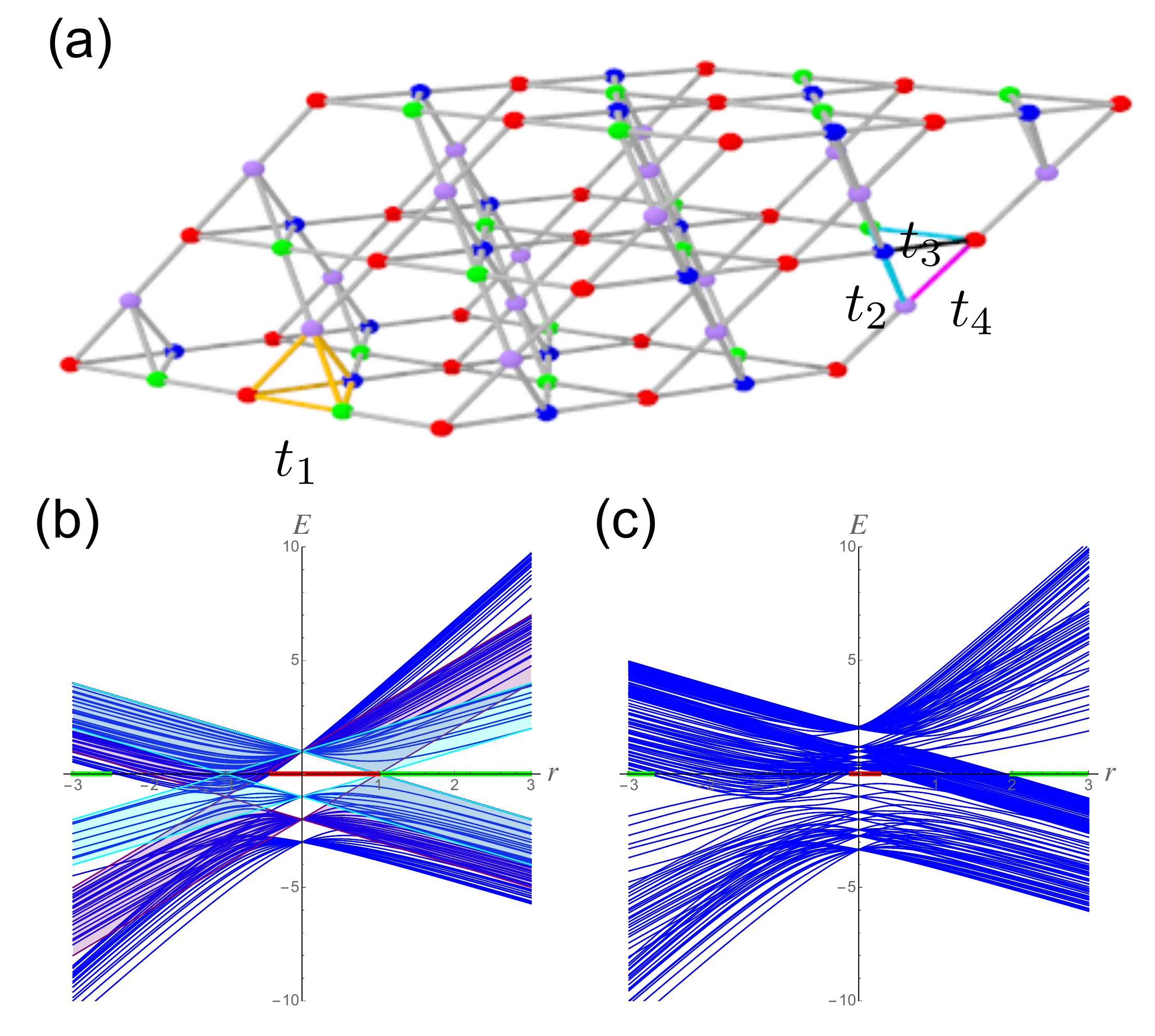}}
\caption{(a) Breathing pyrochlore lattice with the lattices $A$, $B$, $B'$, and $B''$ shown in red, green, blue, and lilac, respectively. The orange, blue, black, and pink bonds have hopping parameters $t_1$, $t_2$, $t_3$, and $t_4$, respectively. (b), (c) The band spectrum for $M = M' = 4$ and (b) $r = r' = r''$ and (c) $r = 2 r' = r''/2$. In (b), the purple and cyan shaded areas correspond to bulk bands originating from the surfaces and hinges of the breathing pyrochlore lattice whereas the rest of the bulk bands originate from the bulk.
}
\label{fig_pyrochlore_lattice_corners}
\end{figure}

{\it Breathing pyrochlore corner states.} We now add a third intermediate lattice, $B''$, to construct the three-dimensional breathing pyrochlore lattice \cite{ezawapap} shown in Fig.~\ref{fig_pyrochlore_lattice_corners}(a). We again assume a real nearest-neighbor hopping model where the up-tetrahedra have hopping strengths $t_1$ whereas the down-tetrahedra have bonds with hopping strengths $t_2$, $t_3$, and $t_4$ as shown in Fig.~\ref{fig_pyrochlore_lattice_corners}(a). The wave-function solution for the corner state is given in Eq.~(\ref{eqexactsolgeneral}) and localizes to either of the eight corners depending on the values of $r =-t_1/t_2$, $r' =-t_1/t_3$, and $r'' =-t_1/t_4$. Energy spectra are shown in Figs.~\ref{fig_pyrochlore_lattice_corners}(b) and \ref{fig_pyrochlore_lattice_corners}(c), where the zero-energy mode localizes to the localization sector $\Gamma_{1,1,1}$ and $\Gamma_{M,M',M''}$ in the red and green region, respectively; isolated zero-energy modes are only found in these two localization sectors, which are again rooted in the properties of the underlying SSH physics. 

\begin{figure}[h]
  \centering
  {\includegraphics[width=1\linewidth]{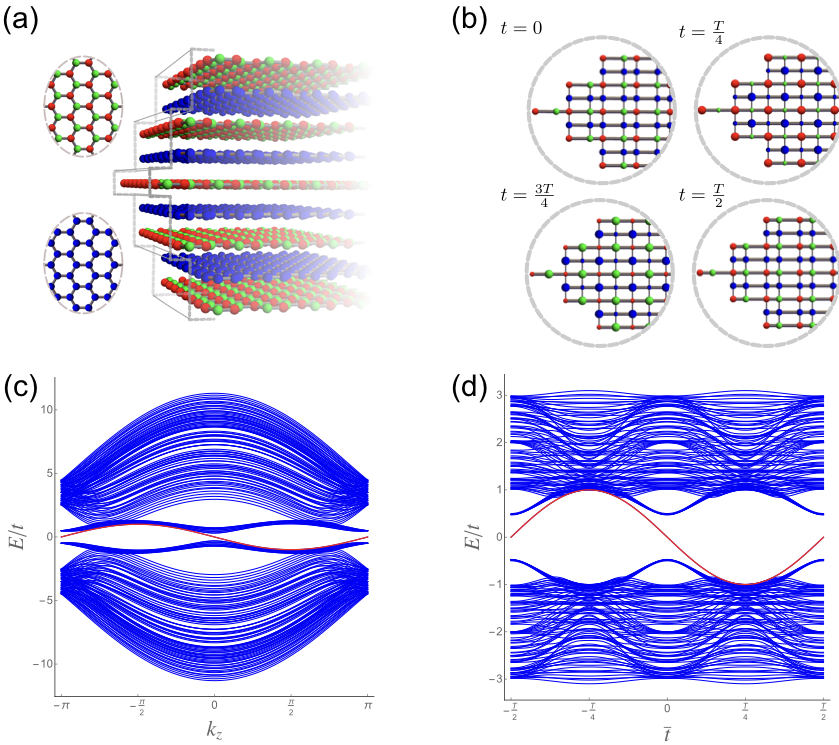}}
\caption{(a) Stacked honeycomb lattices and (b) stacked Rice-Mele chains with the lattices $A$, $B$, and $B'$ in red, green, and blue, respectively, with the corresponding band spectra in (c) with $t=3.5$, $\Delta=-0.5$, and $s=0.5$, and (d) with $t=1$, $s=0.5$, $\Delta(\bar{t})=\pm\sin(2\pi \bar{t}/T)$, and $\delta(\bar{t})=\cos(2\pi \bar{t}/T)$, respectively. In (b), the size of the spheres and the bond lengths correspond to the on-site energy and the hopping strengths which vary as a function of $\bar{t}$. The red bands in (c) and (d) correspond to the chiral hinge states, whereas the closely packed blue bands around the hinge states and the wider packed blue bands originate from the surfaces and bulk, respectively.}
\label{fig_chiral_hinge_states}
\end{figure}

{\it Chiral hinge states.}
Finally, we apply our method to higher-order topological insulators with robust and exactly solvable hinge states and consider explicitly mirror-symmetric 3D insulators. These states of matter are characterized by a mirror Chern number. For simplicity, let us consider a 3D system that is mirror symmetric in the $y$ direction. In this case, the mirror Chern number, $C_m(k_y)$ with $k_y=0,\pi$, is defined as $$C_m(k_y)=\frac{C_+(k_y)-C_-(k_y)}{2},$$ where $C_{+(-)}$ denotes the Chern number of the even (odd) sector, respectively \cite{ChiuYaoRyu,ShiozakiSato,TeoFuKane,MorimotoFurusaki}. For $k_y=0,\pi$, the even (odd) sector is composed of the occupied Bloch waves $|\Psi(k_x,k_y,k_z)\rangle$ that have an eigenvalue $+1(-1)$ under reflection in the $y$ direction. If and only if $C_m(0)+C_m(\pi)=1 \, \textrm{mod} \, 2$, does the system exhibit an odd number of chiral edge states (Dirac cones) on each hinge (surface), that is symmetric under $y\rightarrow -y$ \cite{langhehnpentrifuoppenbrouwer,schindlercookvergnio}. To construct such a model, we stack 2D Chern insulators, which are periodic in $z$ and have edges in the $x$ direction, and stack them in the $y$ direction in a mirror-symmetric way in accordance with the schematic lattice structure in Fig.~\ref{fig_general_structure}(c). In addition, we require that the planes have alternating Chern numbers $\pm 1$ such that the edge modes between two neighboring Chern insulators hybridize on the surface while the edge modes of the layer $m' = (M' +1)/2$ survive. Even though the ordinary Chern numbers $C(0)$ and $C(\pi)$ are identically zero, we find that $C_m(0)+C_m(\pi)=1 \ \textrm{mod }2$. We thus expect the presence of chiral hinge states on the layer $m' = (M' +1)/2$. First, we consider the case in which each 2D Chern insulator corresponds to a honeycomb lattice, as shown schematically in Fig.~\ref{fig_chiral_hinge_states}(a) with the nearest-neighbor hopping parameter $t$ and complex next-nearest-neighbor hopping $\pm i\Delta$ in the $z$ direction, where the sign is positive (negative) if the hopping is (anti)clockwise \cite{haldane}. To ensure that the stacks have alternating Chern numbers, we simply alternate the sign of $\Delta$. Finally, we hybridize the stacks through nearest-neighbor hopping $s$ in the $y$ direction. The dispersion relation as a function of $k_z$ is shown in Fig.~\ref{fig_chiral_hinge_states}(c), and one can distinguish three regimes in the band structure: the bulk bands, the surface bands, and the states living on the hinges (shown in red). The hinge-state dispersion and wave function can be obtained exactly. Indeed, the exact edge-state wave function for an isolated Chern insulator labeled by $m'$ in Fig.~\ref{fig_general_structure}(c) with $M$ sublattices $A$ in the $x$ direction reads
\begin{equation}
\ket{\Psi (k_z)}_{m'}= \sum_{m}\left(\frac{-1}{1+{\rm e}^{i k_z}}\right)^{m} c^\dagger_{A,{m,m'}} (k_z)\ket{0}, \label{eqextrasolutionond}
\end{equation}
with the associated eigenvalue $E(k_z)=2\Delta\sin(k_z)$ originating from the Hamiltonian on the $A$ sublattices. Since each $B'$ stack is sandwiched between two $AB$ stacks, we can let the two exact edge-state wave functions of the $AB$ stacks [Eq.~(\ref{eqextrasolutionond})] interfere destructively on $B'$ by alternating the sign of the wave function such that the wave function of the hinge state is given by 
\begin{equation}
\ket{\Psi_{\rm hinge}(k_z)} = \mathcal{N} \sum_{m'} (-1)^{m'} \ket{\Psi (k_z)}_{m'}.
\end{equation}
This state decays exponentially in the $x$ direction dictated by $r(k_z)=\frac{-1}{1+e^{i k_z}}$, whereas in the $y$ direction we find $r'=-1$ for all $m'$. Hence, for the geometry shown in Fig.~\ref{fig_general_structure}(c), we find that this state localizes on the hinges $\{m,m'\} = \{1, (M'+1)/2\}$ or $\{m,m'\} = \{M, (M'+1)/2\}$ when $|r_1(k_z)|<1$ or $|r_1(k_z)|>1$, respectively. We find $|r_1(k_z)|=1$ at $k_z = \pm 2 \pi/3$, indicating that the hinge state switches localization sectors. Indeed, it can be seen in Fig.~\ref{fig_chiral_hinge_states}(c) that for $k_z=+2\pi/3$ the hinge state attaches to the valence bands, whereas for $k_z=-2\pi/3$ it attaches to the conduction band, such that the state that we have found indeed corresponds to the chiral hinge states.

The model that we introduced above may be difficult to implement experimentally. As an alternative we also consider a truly 2D stack of 1D chains, and simulate the third dimension by adiabatic periodic driving, i.e., time now plays the role of momentum in the $z$ direction. The design strongly resembles the stacking of the Chern insulators and is shown schematically in Fig.~\ref{fig_chiral_hinge_states}(b), and exact hinge states thereof are readily obtained. In this case we stack 1D charge pumps \cite{tknn,Thouless2} with alternating Chern numbers. Each 1D chain corresponds to a two-band Rice-Mele model \cite{ricemele}, where the electrons experience a staggered sublattice potential $\pm \Delta(\bar{t})$, and the hopping strength alternates between $t\pm\delta(\bar{t})$. Here, we consider the case with $\Delta(\bar{t})=-(+)\sin(2\pi \bar{t}/T)$ and $\delta(\bar{t})=\cos(2\pi \bar{t}/T)$ for the $AB$ ($B'$) chains, with $T$ the period of the driving. We hybridize the $AB$ and $B'$ chains by including nearest-neighbor hoppings $s$ in the $y$ direction, as shown schematically in Fig.~\ref{fig_general_structure}(c). Snapshots of the Hamiltonian are schematically depicted in Fig.~\ref{fig_chiral_hinge_states}(b) and the evolution of the band spectrum as a function of $\bar{t}$ is shown in Fig.~\ref{fig_chiral_hinge_states}(d). Note that $\bar{t}$ changes adiabatically slow such that the band spectrum is obtained instantaneously. Again, we find that there are precisely two states crossing the band gap with $r(\bar{t}) = - [t + \delta(\bar{t})]/[t-\delta(\bar{t})]$ and $r' = -1$. The existence of these states is due to the pumping of one electron from the left to the right corner. The spectral flow is stable as long as the mirror symmetry is preserved, where one does not require a mirror-symmetric corner, and only the two edges that intersect at the corner must be related by this mirror symmetry.

{\it Conclusion.} In this Rapid Communciation, we have provided a simple and generic recipe for engineering $D$-dimensional models with exactly solvable eigenstates that exponentially localize to $d$-dimensional boundary motifs, i.e., to surfaces, edges, hinges, corners, etc. Remarkably, these exact solutions are valid for generic tight-binding parameters, providing microscopic insight and an intuition for the phase diagram of these models. We have also shown that this facilitates the understanding of the subtle interplay between symmetries and topology in the context of the recently discovered higher-order topological states of matter, and we hope that the striking simplicity of our construction will be helpful in designing future experimental realizations thereof.

{\it Acknowledgments.}
F.K.K and E.J.B. are supported by the Swedish research council (VR) and the Wallenberg Academy Fellows program of the Knut and Alice Wallenberg Foundation. This work is part of the research programme of the Foundation for Fundamental Research on Matter (FOM), which is part of the Netherlands Organization for Scientific Research (NWO). 









\begin{thebibliography}{10}

\bibitem{tknn}
D.J. Thouless, M. Kohmoto, M.P. Nightingale, and M. den Nijs, {\em Quantized Hall Conduc- tance in a Two-Dimensional Periodic Potential}, \href{https://journals.aps.org/prl/abstract/10.1103/PhysRevLett.49.405}{Phys. Rev. Lett. 49, 405 (1982)}.

\bibitem{haldane}
F.D.M. Haldane, {\em Model for a Quantum Hall Effect without Landau Levels: Condensed-Matter Realization of the "Parity Anomaly"}, \href{http://link.aps.org/doi/10.1103/PhysRevLett.61.2015}{Phys. Rev. Lett. {\bf 61}, 2015 (1988)}.

\bibitem{changzhangfengshenzhang}
C.-Z. Chang, J. Zhang, X. Feng, J. Shen, Z. Zhang, M. Guo, K. Li, Y. Ou, P. Wei, L.-L. Wang, Z.-Q. Ji, Y. Feng, S. Ji, X. Chen, J. Jia, X. Dai, Z. Fang, S.-C. Zhang, K. He, Y. Wang, L. Lu, X.-C. Ma, and Q.-K Xue, {\em Experimental Observation of the Quantum Anomalous Hall Effect in a Magnetic Topological Insulator}, \href{http://science.sciencemag.org/content/340/6129/167}{Science {\bf 340}, 167 (2013)}.

\bibitem{jotzumesserdesbuquoislebratuehlingergreifesslinger}
G. Jotzu, M. Messer, R. Desbuquois, M. Lebrat, T. Uehlinger, D. Greif, and T. Essliner, {\em Experimental realization of the topological Haldane model with ultracold fermions}, \href{http://www.nature.com/nature/journal/v515/n7526/full/nature13915.html}{Nature {\bf 515}, 237 (2014)}.

\bibitem{kaneandmele}
C.L. Kane and E.J. Mele, {\em Quantum Spin Hall Effect in Graphene}, \href{http://journals.aps.org/prl/abstract/10.1103/PhysRevLett.95.226801}{Phys. Rev. Lett. {\bf 95}, 226801 (2005)}.

\bibitem{kaneandmeletwo}
C. L. Kane and E. J. Mele, {\em Z$_2$ Topological Order and the Quantum Spin Hall Effect}, \href{http://journals.aps.org/prl/abstract/10.1103/PhysRevLett.95.146802}{Phys. Rev. Lett. {\bf 95}, 146802 (2005)}.

\bibitem{bernevigzhang}
B. A. Bernevig and S.-C. Zhang, {\em Quantum Spin Hall Effect}, \href{https://journals.aps.org/prl/abstract/10.1103/PhysRevLett.96.106802}{Phys. Rev. Lett. {\bf 96}, 106802 (2006)}.

\bibitem{bernevighugheszhang}
B. A. Bernevig, T. L. Hughes, and S.-C. Zhang, {\em Quantum Spin Hall Effect and Topological Phase Transition in HgTe Quantum Wells}, \href{http://science.sciencemag.org/content/314/5806/1757.long}{Science {\bf 314}, 1757 (2006)}.

\bibitem{koenigwiedmannbruenerothbuhmannmolenkamp}
M. K\"onig, S. Wiedmann, C. Br\"une, A. Roth, H. Buhmann, and L. W. Molenkamp, {\em Quantum Spin Hall Insulator State in HgTe Quantum Wells}, \href{http://science.sciencemag.org/content/318/5851/766.long}{Science {\bf 318}, 766 (2007)}.

\bibitem{klitzingdordapepper}
K. von Klitzing, G. Dorda, and M. Pepper, {\em New Method for High-Accuracy Determination of the Fine-Structure Constant Based on Quantized Hall Resistance}, \href{http://journals.aps.org/prl/abstract/10.1103/PhysRevLett.45.494}{Phys. Rev. Lett. {\bf 45}, 494 (1980)}.

\bibitem{volovik}
G.E. Volovik, {\em The Universe in a Helium Droplet} (Clarendon, Oxford, 2003).

\bibitem{murakami}
S. Murakami, {\em Phase transition between the quantum spin Hall and insulator phases in 3D: emergence of a topological gapless phase}, \href{http://iopscience.iop.org/1367-2630/9/9/356/}{New J. Phys. {\bf 9}, 356 (2007)}.

\bibitem{wanturnerbishwanathsavrasov}
X. Wan, A. M. Turner, A. Vishwanath, and S. Y. Savrasov, {\em Topological semimetal and Fermi-arc surface states in the electronic structure of pyrochlore iridates}, \href{http://journals.aps.org/prb/abstract/10.1103/PhysRevB.83.205101}{Phys. Rev. B {\bf 83}, 205101 (2011)}.

\bibitem{burkovbalents}
A.A. Burkov and L. Balents, {\em Weyl Semimetal in a Topological Insulator Multilayer}, \href{http://journals.aps.org/prl/abstract/10.1103/PhysRevLett.107.127205}{Phys. Rev. Lett. {\bf 107}, 127205 (2011)}.

\bibitem{luwangyeranfujoannopoulossoljacic}
L. Lu, Z. Wang, D. Ye, L. Ran, L. Fu, J.D. Joannopoulos, and M. Solja\v{c}i\'{c}, {\em Experimental observation of Weyl points}, \href{http://science.sciencemag.org/content/349/6248/622.abstract?ijkey=FW/d7hvxzlHnA&keytype=ref&siteid=sci}{Science {\bf 349}, 622 (2015)}.

\bibitem{xubelopolskialidoustetal}
S.-Y. Xu, I. Belopolski, N. Alidoust, M. Neupane, G. Bian, C. Zhang, R. Sankar, G. Chang, Z. Yuan, C.-C. Lee, S.-M. Huang, H. Zheng, J. Ma, D.S. Sanchez, B.K. Wang, A. Bansil, F. Chou, P.P. Shibayev, H. Lin, S. Jia, and M.Z. Hasan, {\em Discovery of a Weyl fermion semimetal and topological Fermi arcs}, \href{http://science.sciencemag.org/content/349/6248/613}{Science {\bf 349}, 613 (2015)}.

\bibitem{lvwengwantmiaoetal}
B.-Q. Lv, H.-M. Weng, B.-B. Fu, X.-P. Wang, H. Miao, J. Ma, P. Richard, X.-C. Huang, L.-X. Zhao, G.-F. Chen, Z. Fang, X. Dai, T. Qian, and H. Ding, {\em Experimental Discovery of Weyl Semimetal TaAs}, \href{http://journals.aps.org/prx/abstract/10.1103/PhysRevX.5.031013}{Phys. Rev. X {\bf 5}, 031013 (2015)}.

\bibitem{benalcazarbernevighughes}
W.A. Benalcazar, B.A. Bernevig, and T.L. Hughes, {\em Quantized electric multipole insulators}, \href{http://science.sciencemag.org/content/357/6346/61}{Science {\bf 357}, 61 (2017)}.

\bibitem{langhehnpentrifuoppenbrouwer}
J. Langbehn, Y. Peng, L. Trifunovic, F. von Oppen, P.W. Brouwer, {\em Reflection symmetric second-order topological insulators and superconductors}, \href{https://journals.aps.org/prl/abstract/10.1103/PhysRevLett.119.246401}{Phys. Rev. Lett. {\bf 119}, 246401 (2017)}.

\bibitem{linhughes}
M. Lin and T. Hughes, {\em Topological Quadrupolar Semimetals}, \href{https://arxiv.org/abs/1708.08457}{arXiv:1708.08457 (2017)}.

\bibitem{songfangfang}
Z. Song, Z. Fang, and C. Fang, {\em $(d-2)$-dimensional edge states of rotation symmetry protected topological states}, \href{https://journals.aps.org/prl/abstract/10.1103/PhysRevLett.119.246402}{Phys. Rev. Lett. {\bf 119}, 246402 (2017)}.

\bibitem{benalcazarbernevighughesagain}
W.A. Benalcazar, B.A. Bernevig, T.L. Hughes, {\em Electric Multipole Moments, Topological Multipole Moment Pumping, and Chiral Hinge States in Crystalline Insulators}, \href{https://arxiv.org/abs/1708.04230}{arXiv:1708.04230 (2017)}.

\bibitem{schindlercookvergnio}
F. Schindler, A.M. Cook, M.G. Vergniory, Z. Wang, S.S.P. Parkin, B.A. Bernevig, and T. Neupert, {\em Higher-Order Topological Insulators}, \href{https://arxiv.org/abs/1708.03636}{arXiv:1708.03636}.

\bibitem{ezawapap}
M. Ezawa, {\em Higher-order topological insulators and semimetals on the breathing Kagome and pyrochlore lattices}, \href{https://arxiv.org/abs/1709.08425}{arXiv: 1709.08425 (2017)}.

\bibitem{xuxuewan}
Y. Xu, R. Xue, S. Wan, {\em Topological Corner States on Kagome Lattice Based Chiral Higher-Order Topological Insulator}, \href{https://arxiv.org/abs/1711.09202}{arXiv: 1711.09202 (2017)}.

\bibitem{parameswaranwan}
S.A. Parameswaran and Y. Wan, {\em Viewpoint: Topological Insulators Turn a Corner}, \href{https://physics.aps.org/articles/v10/132}{Physics {\bf 10}, 132 (2017)}.

\bibitem{imhofbergerbayerbrehmmolenkampkiesslingschindlerleegreiterneupertthomale}
S. Imhof, C. Berger, F. Bayer, J. Brehm, L. Molenkamp, T. Kiessling, F. Schindler, C.H. Lee, M. Greiter, T. Neupert, and R. Thomale, {\em Topolectrical circuit realization of topological corner modes}, \href{https://arxiv.org/abs/1708.03647}{arXiv:1708.03647 (2017)}.

\bibitem{garciaperissstrunkbilallarsenvillanuevahuber}
M. Serra-Garcia, V. Peri, R. S\"{u}sstrunk, O.R. Bilal, T. Larsen, L.G. Villanueva, and S.D. Huber, {\em Observation of a phononic quadrupole topological insulator}, \href{https://arxiv.org/abs/1708.05015}{arXiv:1708.05015 (2017)}.

\bibitem{petersonbenalcazarhughesbahl}
C.W. Peterson, W.A. Benalcazar, T.L. Hughes, and G. Bahl, {\em Demonstration of a quantized microwave quadrupole insulator with topologically protected corner states}, \href{https://arxiv.org/abs/1710.03231}{arXiv:1710.03231 (2017)}.

\bibitem{kunsttrescherbergholtz}
F.K. Kunst, M. Trescher, and E.J. Bergholtz, {\em Anatomy of topological surface states: Exact solutions from destructive interference on frustrated lattices}, \href{https://journals.aps.org/prb/abstract/10.1103/PhysRevB.96.085443}{Phys. Rev. B {\bf 96}, 085443 (2017)}.

\bibitem{bergholtz2015}
E. J. Bergholtz, Z. Liu, M. Trescher, R. Moessner and M. Udagawa,
{\em Topology and Interactions in a Frustrated Slab: Tuning from Weyl Semimetals to $\mathcal{C}>1$ Fractional Chern Insulators},
\href{http://link.aps.org/doi/10.1103/PhysRevLett.114.016806}{Phys. Rev. Lett. {\bf 114}, 016806 (2015)}.

\bibitem{ssh}
W.P. Su, J.R. Schrieffer, and A.J. Heeger, {\em Solitons in Polyacetylene}, \href{https://journals.aps.org/prl/abstract/10.1103/PhysRevLett.42.1698}{Phys. Rev. Lett. {\bf 42}, 1698 (1979)}.

\bibitem{ChiuYaoRyu}
C.-K. Chiu, H. Yao, and S. Ryu, 
{\em Classification of topological Insulators and superconductors in the presence of reflection symmetry},
\href{https://link.aps.org/doi/10.1103/PhysRevB.88.075142}{Phys. Rev. B {\bf 88}, 075142 (2013)}.

\bibitem{ShiozakiSato}
K. Shiozaki and M. Sato
{\em Topology of crystalline insulators and superconductors},
\href{https://link.aps.org/doi/10.1103/PhysRevB.90.165114}{ Phys. Rev. B {\bf 90}, 165114 (2014)}.

\bibitem{TeoFuKane}
J.C.Y. Teo, L. Fu, and C. L. Kane
{\em Surface states and topological invariants in three-dimensional topological insulators: Application to ${\text{Bi}}_{1\ensuremath{-}x}{\text{Sb}}_{x}$},
\href{https://link.aps.org/doi/10.1103/PhysRevB.78.045426}{Phys. Rev. B {\bf 78}, 045426 (2008)}.

\bibitem{MorimotoFurusaki}
T.Morimoto and A. Furusaki
{\em Topological classification with additional symmetries from Clifford algebras},
\href{https://link.aps.org/doi/10.1103/PhysRevB.88.125129}{Phys. Rev. B {\bf 88}, 125129 (2013)}.

\bibitem{Thouless2}
D. J. Thouless
{\em Quantization of Particle Transport}. \href{https://link.aps.org/doi/10.1103/PhysRevB.27.6083} {Phys. Rev. B {\bf 27}, 6083 (1983)}.

\bibitem{ricemele}
M. J. Rice and E. J. Mele,
{\em Elementary Excitations of a Linearly Conjugated Diatomic Polymer},
\href{https://doi.org/10.1103/PhysRevLett.49.1455}{Phys. Rev. Lett. {\bf 49}, 1455 (1982)}.

\end{thebibliography}
\end{document}